\documentclass[10pt,twocolumn,twoside]{IEEEtran}

\usepackage{amsmath,amssymb,amsfonts,mathtools,dsfont,multirow,relsize}
\usepackage{algorithm2e}
\usepackage{cite}
\usepackage{color}
\usepackage{caption,subcaption}
\usepackage{float}
\usepackage{hyperref}

\newcommand{\sfB}{\mathsf{B}}

\newcommand{\sfD}{\mathsf{D}}

\newcommand{\sfP}{\mathsf{P}}

\newcommand{\bp}{\boldsymbol{p}}

\newcommand{\bv}{\boldsymbol{v}}
\newcommand{\by}{\boldsymbol{y}}

\newcommand{\bP}{\boldsymbol{P}}

\newcommand{\bY}{\boldsymbol{Y}}

\newcommand{\calH}{\mathcal{H}}

\newcommand{\bbE}{\mathbb{E}}

\newcommand{\bbR}{\mathbb{R}}

\newcommand{\bdelta}{\boldsymbol{\delta}}

\newcommand{\defeq}{\vcentcolon=}

\DeclareMathOperator*{\argmin}{argmin}

\begin{document}
\renewcommand{\thepage}{}

\title{{Optimal Decision Rules for Simple Hypothesis Testing under General Criterion Involving Error Probabilities}}

\author{Berkan Dulek\thanks{B. Dulek is with the Department of Electrical and Electronics Engineering, Hacettepe University, Beytepe Campus, Ankara 06800, Turkey,
{\tt e-mail: berkan@ee.hacettepe.edu.tr}.}, Cuneyd Ozturk, \textit{Student Member, IEEE}, and Sinan Gezici, \emph{Senior Member, IEEE}\thanks{C. Ozturk and S. Gezici are with the Department of Electrical and Electronics Engineering, Bilkent University, Ankara 06800, Turkey, {\tt e-mails: \{cuneyd,gezici\}@ee.bilkent.edu.tr}.}}


\maketitle

\pagenumbering{arabic}


\begin{abstract}
The problem of simple $M-$ary  hypothesis testing under a generic performance criterion that depends on arbitrary functions of error probabilities is considered. Using results from convex analysis, it is proved that an optimal decision rule can be characterized as a randomization among \emph{at most} two deterministic decision rules, of the form reminiscent to Bayes rule, if the boundary points corresponding to each rule have zero probability under each hypothesis. Otherwise, a randomization among \emph{at most} $M(M-1)+1$ deterministic decision rules is sufficient. The form of the deterministic decision rules are explicitly specified. Likelihood ratios are shown to be sufficient statistics. Classical performance measures including Bayesian, minimax, Neyman-Pearson, generalized Neyman-Pearson, restricted Bayesian, and prospect theory based approaches are all covered under the proposed formulation. A numerical example is presented for prospect theory based binary hypothesis testing.

\end{abstract}

\section{Problem Statement} \label{sec:pre}

Consider a detection problem with $M$ simple hypotheses:
\begin{equation} \label{eq:ht}
\calH_j:  \bY \thicksim f_j(\cdot),\text{ with } j=0,1,\ldots,M-1,
\end{equation}
where the random observation $\bY$ takes values from an observation set $\Gamma$ with $\Gamma\subset{\mathbb{R}}^N$. Depending on whether the observed random vector $\bY \in \Gamma$ is continuous-valued or discrete-valued, $f_j(\by)$ denotes either the probability density function (pdf) or the probability mass function (pmf) under hypothesis $\calH_j$. For compactness of notation, the term \emph{density} is used for both pdf and pmf. In order to decide among the hypotheses, we consider the set of pointwise randomized decision functions, denoted by $\sfD$, i.e.,  $\bdelta\defeq (\delta_0,\delta_1,\ldots,\delta_{M-1})\in \sfD$ such that $\sum_{i=0}^{M-1} \delta_i(\by)=1$ and $\delta_i(\by) \in [0,1]$ for  $0\leq i\leq M-1$ and $\by \in \Gamma$. More explicitly, given the observation $\by$, the detector decides in favor of hypothesis $\calH_i$ with probability $\delta_i(\by)$. Then, the probability of choosing hypothesis $\calH_i$ when hypothesis $\calH_j$ is true, denoted by $p_{ij}$ with $0\leq i,j\leq M-1$, is given by
\begin{equation} \label{eq:pij}
p_{ij} \defeq  \bbE_{j}[\delta_i(\by)]= \int_{\Gamma} \delta_i(\by) f_j(\by)\mu(d\by),
\end{equation}
where $\bbE_j[\cdot]$ denotes expected value under hypothesis $\calH_j$ and $\mu(d\by)$ is used in \eqref{eq:pij} to denote the $N-$fold integral and sum for continuous and discrete cases, respectively. Let $\bp(\bdelta)$ denote the (column) vector containing all pairwise error probabilities $p_{ij}$ for $0\leq i,j\leq M-1$ and $i\neq j$ corresponding to the decision rule $\bdelta$. It is sufficient to include only the pairwise error probabilities in $\bp(\bdelta)$, i.e., $p_{ij}$ with $i\neq j$. To see this, note that  \eqref{eq:pij} in conjunction with $\sum_{i=0}^{M-1} \delta_i(\by)=1$ imply $\sum_{i=0}^{M-1} p_{ij}=1$, from which we get the probability of correctly identifying hypothesis $\calH_i$ as $p_{ii}=1-\sum_{i=0, i\neq j}^{M-1} p_{ij}$.

For $M$-ary hypothesis testing, we consider a generic decision criterion that can be expressed in terms of the error probabilities as follows:
\begin{align}
\underset{\bdelta \in \sfD}{\textrm{minimize}}\qquad\ & g_0(\bp(\bdelta))\nonumber\\
\textrm{subject to}\qquad & g_i(\bp(\bdelta)) \leq 0, \quad i=1,2,\ldots,m \nonumber\\
& h_j(\bp(\bdelta)) = 0, \quad j=1,2,\ldots,p \label{eq:opt1}
\end{align}
where $g_i$ and $h_j$ denote arbitrary functions of the pairwise error probability vector. Classical hypothesis testing criteria such as Bayesian, minimax, Neyman-Pearson (NP) \cite{DETEST_Poor}, generalized Neyman-Pearson \cite{PROB_Lehmann}, restricted Bayesian \cite{Hodges52}, and prospect theory based hypothesis testing \cite{Gezici2018} are all special cases of the formulation in \eqref{eq:opt1}. For example, in the restricted Bayesian framework, the Bayes risk with respect to (w.r.t.) a certain prior is minimized subject to a constraint on the maximum conditional risk \cite{Hodges52}:
\begin{align}\label{eq:opt1ex5}
\underset{\bdelta \in \sfD}{\textrm{minimize}}\quad &\quad\, r_B(\bdelta)\nonumber\\
\textrm{subject to}\quad & \underset{0\leq j \leq M-1}{\textrm{max}}\  R_j(\bdelta)\ \leq \alpha
\end{align}
for some $\alpha\geq \alpha_m$, where $\alpha_m$ is the maximum conditional risk of the minimax procedure \cite{DETEST_Poor}. The conditional risk when hypothesis $H_j$ is true, denoted by $R_j(\bdelta)$, is given by $R_j(\bdelta) =  \sum_{i=0}^{M-1} c_{ij} p_{ij}$
and the Bayes risk is expressed as $r_B(\bdelta) = \sum_{j=0}^{M-1} \pi_j R_j(\bdelta)$,  where $\pi_j$ denotes the \emph{a priori} probability of hypothesis $\calH_j$ and $c_{ij}$ is the cost incurred by choosing hypothesis $\calH_i$ when in fact hypothesis $\calH_j$ is true. Hence, \eqref{eq:opt1ex5} is a special case of \eqref{eq:opt1}.

In this letter, for the first time in the literature, we provide a unified characterization of optimal decision rules for simple hypothesis testing under a general criterion involving error probabilities.

\section{Preliminaries} \label{sec:pre}

Let $\bv$ be a real (column) vector of length $M(M-1)$ whose elements are denoted as $v_{ij}$ for $0\leq i,j\leq M-1$ and $i\neq j$. Next, we present an optimal deterministic decision rule that minimizes the weighted sum of $p_{ij}$'s with \emph{arbitrary real} weights $\bv$.\footnote{In classical Bayesian $M-$ary hypothesis testing, $v_{ij}=\pi_j (c_{ij}-c_{jj})$.}

\subsection{Optimal decision rule that minimizes $\bv^T \bp(\bdelta)$}

The corresponding weighted sum of pairwise error probabilities can be written as
\begin{align}\label{eq:lemmapf}
  \bv^T \bp(\bdelta) &= \sum_{i=0}^{M-1}\sum_{j=0, j\neq i}^{M-1} v_{ij} p_{ij} \nonumber \\
  &=  \int_{\Gamma}  \sum_{i=0}^{M-1} \delta_i(\by) \left(\sum_{j=0, j\neq i}^{M-1} v_{ij}  f_j(\by)\right)\mu(d\by),
\end{align}
where  \eqref{eq:pij} is substituted for $p_{ij}$ in \eqref{eq:lemmapf}. Defining
$V_i(\by)\defeq \sum_{j=0, j\neq i}^{M-1} v_{ij}  f_j(\by)$,
we get
\begin{align}\label{eq:lemmapf2}
  \bv^T \bp(\bdelta) &= \int_{\Gamma}  \sum_{i=0}^{M-1} \delta_i(\by) V_i(\by)\, \mu(d\by) \nonumber\\
  &\geq \int_{\Gamma}  \min_{0\leq i \leq M-1}\{ V_i(\by)\}\ \mu(d\by)
\end{align}
The lower bound in \eqref{eq:lemmapf2} is achieved if, for all $\by\in \Gamma$, we set
\begin{equation}\label{eq:bayesrule}
\delta_{\ell}(\by)=1 \text{ for } \ell=\argmin_{0\leq i \leq M-1} V_i(\by)
\end{equation}
(and hence, $\delta_{i}(\by)=0$ for all $i\neq \ell$), i.e., each observed vector $\by$ is assigned to the corresponding hypothesis that minimizes $V_i(\by)$ over all $0\leq i \leq M-1$. In case where there are multiple hypotheses that achieve the same minimum value of $V_{\ell}(\by)$  for a given observation $\by$, the ties can be broken by arbitrarily selecting one of them since the boundary decision does not affect the decision criterion $\bv^T \bp(\bdelta)$. However, pairwise probabilities for erroneously selecting hypotheses $\calH_i$ and $\calH_j$ will change if the set of boundary points
\begin{multline}\label{eq:gammaij}
\sfB_{i,j}(\bv)\defeq \{\by\in\Gamma\, :\, V_i(\by)=V_j(\by)\leq V_k(\by)\\ \text{ for all } 0\leq k\leq M-1 ,k\neq i,k\neq j\}
\end{multline}
occurs with nonzero probability. We also define the set of all boundary points
\begin{equation}\label{eq:bound}
  \sfB(\bv)\defeq \underset{ \substack{0\leq i\leq M-1\\ i< j\leq M-1}}\bigcup \sfB_{i,j}(\bv)
\end{equation}
and the complimentary set where $V_{i}(\by)$ for some $0\leq i \leq M-1$ is strictly smaller than the rest:
\begin{multline}\label{eq:bound2}
  \bar{\sfB}(\bv)\defeq \Gamma\setminus \sfB(\bv) = \{\by\in\Gamma\, :\, V_i(\by)< V_j(\by), \text{ for some } \\ 0\leq i\leq M-1 \text{ and all } 0\leq j\leq M-1, j\neq i\}
\end{multline}

\subsection{The set of achievable pairwise error probability vectors}

Let $\sfP$ denote the set of all pairwise error probability vectors that can be achieved by randomized decision functions $\bdelta \in \sfD$, i.e., $\sfP\defeq \{\bp(\bdelta)\,:\, \bdelta \in \sfD\}$. In this part, we present some properties of $\sfP$.

\emph{\textbf{Property 1:} $\sfP$ is a convex set.}

\textbf{Proof:} Let $\bp^1(\bdelta^1)$ and $\bp^2(\bdelta^2)$ be two pairwise error probability vectors obtained by employing randomized decision functions $\bdelta^1$ and $\bdelta^2$, respectively. Then, for any $\theta$ with $0\leq \theta\leq 1$, $\bp_{\theta}=\theta \bp^1(\bdelta^1) + (1-\theta) \bp^2(\bdelta^2) \in \sfP$ since $\bp_{\theta}$ is the pairwise error probability vector corresponding to the randomized decision rule $\theta \bdelta^1 + (1-\theta)\bdelta^2 $ as seen from \eqref{eq:pij}.

\emph{\textbf{Property 2:} Let $\bp_0$ be a point on the boundary of $\sfP$. There exists a hyperplane $\{\bp\,:\, \bv^T \bp = \bv^T \bp_0\}$ that is tangent to $\sfP$ at $\bp_0$ and $\bv^T \bp \geq  \bv^T \bp_0$ for all $\bp\in\sfP$.}

\textbf{Proof:} Follows immediately from the supporting hyperplane theorem \cite[Sec.~2.5.2]{CONVEX_Boyd}.

\section{Characterization of  Optimal Decision Rule} \label{sec:res}

In order to characterize the solution of \eqref{eq:opt1}, we first present the following lemma.

\emph{\textbf{Lemma:} Let $\bp_0$ be a point on the boundary of $\sfP$ and $\{\bp\,:\, \bv^T \bp = \bv^T \bp_0\}$ be a supporting hyperplane to $\sfP$ at the point $\bp_0$.\\
Case 1: Any deterministic decision rule of the form given in \eqref{eq:bayesrule} corresponding to the weights specified by $\bv$ yields $\bp_0$ if  $\sfB(\bv)$, defined in \eqref{eq:bound}, has zero probability under all hypotheses.\\
Case 2: $\bp_0$ is achieved by a randomization among at most $M(M-1)$ deterministic decision rules of the form given in \eqref{eq:bayesrule}, all corresponding to the same weights specified by $\bv$, if  $\sfB(\bv)$, defined in \eqref{eq:bound}, has nonzero probability under some hypotheses.}

\textbf{Proof:} See Appendix~\ref{app:lemma}.

It should be noted that the condition in case 1 of the lemma, i.e., $\sfB(\bv)$ has zero probability under all hypotheses,  is not difficult to satisfy. A simple example is when the observation under hypothesis $\calH_i$ is Gaussian distributed with mean $\mu_i$ and variance $\sigma^2$ for all $0\leq i\leq M-1$. Furthermore, the lemma implies that any extreme point of the convex set $\sfP$, i.e., any point on the boundary of the convex set $\sfP$ that is not a convex combination of any other points in the set, can be achieved by a deterministic decision rule of the form \eqref{eq:bayesrule} without any randomization. The points that are on the boundary but not extreme points can be obtained via randomization as stated in case 2.

Next, we present a unified characterization of the optimal decision rule for problems that are in the form of \eqref{eq:opt1}.
We suppose that the problem in \eqref{eq:opt1} is feasible and let $\bdelta^{\ast}$ and $\bp^{\ast}(\bdelta^{\ast})$ denote an  optimal decision rule and the corresponding pairwise error probabilities, respectively.

\emph{\textbf{Theorem:} An optimal decision rule that solves \eqref{eq:opt1} can be obtained as\\
Case 1: a randomization among at most two deterministic decision rules of the form given in \eqref{eq:bayesrule}, each specified by some real $\bv$, if  $\sfB(\bv)$, defined in \eqref{eq:bound}, has zero probability under all hypotheses for all real $\bv$; otherwise\\
Case 2: a randomization among at most $M(M-1)+1$ deterministic decision rules of the form given in \eqref{eq:bayesrule}, one specified by some real $\bv$ and the remaining $M(M-1)$ correspond to the same weights specified by another real $\bv$.}

\textbf{Proof:} If the optimal point $\bp^{\ast}(\bdelta^{\ast})$ is on the boundary of $\sfP$, then the lemma takes care of the proof. Here, we consider the case when  $\bp^{\ast}(\bdelta^{\ast})$ is an interior point of $\sfP$. First, we  pick an arbitrary $\bv^1 \in \bbR^{M(M-1)}$ and derive the optimal deterministic decision rule according to \eqref{eq:bayesrule}. Let $\bp^1$ denote the pairwise error probability vector corresponding to the employed decision rule. Then, we move along the ray that originates from  $\bp^1$ and passes through $\bp^{\ast}(\bdelta^{\ast})$. Since $\bP$ is bounded, this ray will intersect with the boundary of $\bP$ at some point, say $\bp^2$. If the condition in case 1 is satisfied, then by lemma-case 1, there exists a deterministic decision rule of the form given in \eqref{eq:bayesrule} that yields $\bp^2$. Otherwise, by lemma-case 2,  $\bp^2$ is achieved by a randomization among at most $M(M-1)$ deterministic decision rules of the form given in \eqref{eq:bayesrule}, all sharing the same weight vector $\bv^2$. Since $\bp^{\ast}(\bdelta^{\ast})$ resides on the line segment that connects  $\bp^1$ to $\bp^2$, it can be attained by appropriately randomizing among the decision rules that yield $\bp^1$ and $\bp^2$.\hfill$\blacksquare$

When the optimization problem in \eqref{eq:opt1} possesses certain structure, the maximum number of deterministic decision rules required to achieve optimal performance may be reduced below those given in the theorem. For example, suppose that the objective is a concave function of $\bp$ and there are a total of $n$ constraints in \eqref{eq:opt1} which are all linear in $\bp$ (i.e., the feasible set, denoted by $\sfP'$, is the intersection of $\sfP$ with halfspaces and hyperplanes). It is well known that the minimum of a concave function over a closed bounded convex set is achieved at an extreme point \cite{CONVEX_Boyd}. Hence, in this case, the optimal point $\bp^{\ast}$ is an extreme point of $\sfP'$. By Dubin's theorem \cite{Witsen80}, any extreme point of $\sfP'$ can be written as a convex combination of $n+1$ or fewer extreme points of $\sfP$. Since any extreme point of $\sfP$ can be achieved by a deterministic decision rule of the form \eqref{eq:bayesrule}, the optimal decision rule is obtained as a randomization among at most $n+1$  deterministic decision rules of the form \eqref{eq:bayesrule}.
If there are no constraints in \eqref{eq:opt1}, i.e., $n=0$, the deterministic decision rule given in \eqref{eq:bayesrule} is optimal and no randomization is required with a concave objective function.

An immediate and important corollary of the theorem is given below.

\emph{\textbf{Corollary:} Likelihood ratios are sufficient statistics for simple $M-$ary hypothesis testing under any decision criterion that is expressed in terms of arbitrary functions of error probabilities as specified in \eqref{eq:opt1}.}

\textbf{Proof:} It is stated in the theorem that a solution of the generic optimization problem in \eqref{eq:opt1} can be expressed in terms of decision rules of the form given in \eqref{eq:bayesrule}. These decision rules only involve comparisons among $V_i(\by)$'s, which are linear w.r.t. the density terms $f_i(\by)$'s. Normalizing  $f_i(\by)$'s with $f_0(\by)$ and defining $L_i(\by)\defeq f_i(\by)/f_0(\by)$, we see that an optimal decision rule that solves the problem in \eqref{eq:opt1} depends on the observation $\by$ only through the likelihood ratios.\hfill$\blacksquare$

\section{Numerical Examples}\label{sec:Nume}

In this section, numerical examples are presented by considering a binary hypothesis testing problem; i.e., $M=2$ in \eqref{eq:ht}. Suppose that a bit ($0$ or $1$) is sent over two independent binary channels to a decision maker, which aims to make an optimal decision based on the binary channel outputs. The output of binary channel $k$ is denoted by $y_k\in \{0,1\},\ k=1,2$, and the decision maker declares its decision based on $\by=[y_1,y_2]$. The probability that the output of binary channel $k$ is $i$ when bit $j$ is sent is denoted by $p_{ij}^{(k)}$ for $0\leq i,j\leq1$ with $p_{0j}^{(k)}+p_{1j}^{(k)}=1$. Then, the pmf of $\by$ under $\calH_j$ is given by
\begin{align}\label{eq:condPDF}
f_j(\by)=\begin{cases}
p_{0j}^{(1)}p_{0j}^{(2)}\,,&{\textrm{if}}~\by=[0,0]\\
p_{0j}^{(1)}p_{1j}^{(2)}\,,&{\textrm{if}}~\by=[0,1]\\
p_{1j}^{(1)}p_{0j}^{(2)}\,,&{\textrm{if}}~\by=[1,0]\\
p_{1j}^{(1)}p_{1j}^{(2)}\,,&{\textrm{if}}~\by=[1,1]
\end{cases}
\end{align}
for $j\in\{0,1\}$. As in the previous sections, the pairwise error probability vector of the decision maker for a given decision rule $\bdelta$ is represented by $\bp(\bdelta)$, which is expressed as $\bp(\bdelta)=[p_{10},p_{01}]^T$ in this case. It is assumed that the decision maker knows the conditional pdfs in \eqref{eq:condPDF}.

%

In this section, a special case of \eqref{eq:opt1} is considered based on prospect theory by focusing on a behavioral decision maker \cite{Gezici2018,Weight2,Weight1,PTadv}. In particular, 
there exist no constraints (i.e., $m=p=0$ in \eqref{eq:opt1}) and the objective function in \eqref{eq:opt1} is expressed as
\begin{gather}\label{eq:objectiveprospect}
\vspace{-3mm}
g_0(\bp(\bdelta)) = \sum_{i = 0}^{1} \sum_{j = 0}^{1} w(P(\mathcal{H}_i \text{ is selected \& } \mathcal{H}_j \text{ is true} )) v(c_{ij})
\end{gather}

\vspace{-2mm}

\noindent where $w(\cdot)$ is a weight function and $v(\cdot)$ is a value function, which characterize how a behavioral decision maker distorts probabilities and costs, respectively \cite{Gezici2018}, and $P(\cdot)$ denotes the probability of its argument. In the numerical examples, the following weight function is employed:
$w(p)=\frac{p^{\kappa}}{(p^{\kappa}+(1-p)^{\kappa})^{1/\kappa}}$ \cite{Gezici2018,Weight2,Weight1,PTadv}. In addition, the other parameters are set as $v(c_{00}) = 3$, $v(c_{01}) = 10$, $v(c_{10}) = 20$, and $v(c_{11}) = 7$. Furthermore, the prior probabilities of bit $0$ and bit $1$ are assumed to be equal.

\begin{figure}
\vspace{-0.4cm}
	\includegraphics[width=\linewidth]{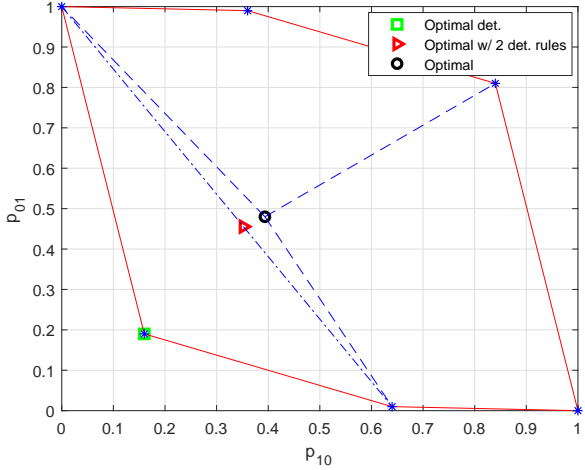}
	\caption{Convex hull of pairwise error probability vectors corresponding to deterministic decision rules in \eqref{eq:bayesrule}, and pairwise error probability vectors corresponding to decision rules which yield the minimum objectives attained via no randomization (marked with square), randomization of two (marked with triangle) and three deterministic decision rules (marked with circle), where $p_{10}^{(1)} = p_{10}^{(2)} = 0.4$, $p_{01}^{(1)} = p_{01}^{(2)} = 0.1$, and  $\kappa=5$.}
	\label{fig:Pd=0.9}
\end{figure}

The aim of the decision maker is to obtain a decision rule that minimizes \eqref{eq:objectiveprospect}. In the first example, $\kappa$ is set to $5$, and the parameters of the binary channels are selected as $p_{10}^{(1)} = p_{10}^{(2)} = 0.4$ and $p_{01}^{(1)} = p_{01}^{(2)} = 0.1$. In this case, it can be shown via \eqref{eq:condPDF} that there exist $6$ different deterministic decision rules in the form of \eqref{eq:bayesrule}, which achieve the pairwise error probability vectors marked with blue stars in Fig.~\ref{fig:Pd=0.9}. The convex hull of these pairwise error probability vectors is also illustrated in the figure. Over these deterministic decision rules (i.e., in the absence of randomization), the minimum achievable value of \eqref{eq:objectiveprospect} becomes $0.1901$, which corresponds to the pairwise error probability vector shown with the green square in Fig.~\ref{fig:Pd=0.9}. If randomization between two deterministic decision rules in the form of \eqref{eq:bayesrule} is considered, the resulting minimum objective value becomes $0.0422$, and the corresponding pairwise error probability vector is indicated with the red triangle in the figure. On the other hand, in compliance with the theorem (case 2), the minimum value of \eqref{eq:objectiveprospect} is achieved via randomization of (at most) three deterministic decision rules in the form of \eqref{eq:bayesrule} (since $M(M-1)+1=3$). In this case, the optimal decision rule randomizes among $\delta_1$, $\delta_2$, and $\delta_3$, with randomization coefficients of $0.41$, $0.51$, and $0.08$, respectively, as given below:
\begin{align}
\delta_1(\by) &=   0~\text{for all~} \by \nonumber \\
\delta_2(\by) &=
 \begin{cases}
  0\,,  \text{ if } \by \in\{[0,1],[1,0],[1,1]\} \\
  1\,, \text{ if }  \by = [0,0]  \label{eq:decrules} \\
\end{cases} \\
\delta_3(\by) &=
\begin{cases}
0\,,  \text{ if } \by = [1,1] \\
1\,, \text{ if }  \by \in\{[0,0],[0,1],[1,0]\} \nonumber
\end{cases}
\end{align}
This optimal decision rule achieves the lowest objective value of $0.0400$, and the corresponding pairwise error probability vector is marked with the black circle in Fig.~\ref{fig:Pd=0.9}. Hence, this example shows that randomization among three deterministic decision rules may be required to obtain the solution of \eqref{eq:opt1}.

\begin{figure}
\vspace{-0.4cm}
	\includegraphics[width=\linewidth]{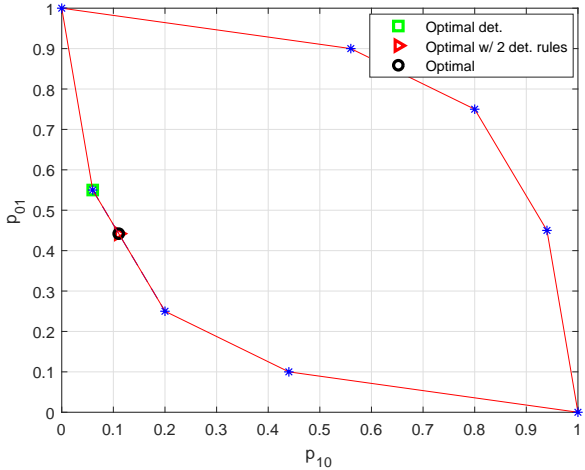}
	\caption{Convex hull of pairwise error probability vectors corresponding to deterministic decision rules in \eqref{eq:bayesrule}, and pairwise error probability vectors corresponding to decision rules which yield the minimum objectives attained via no randomization (marked with square), randomization of two (marked with triangle) and three deterministic decision rules (marked with circle), where $p_{10}^{(1)} = 0.3$,  $p_{10}^{(2)} = 0.2$, $p_{01}^{(1)} = 0.4$, and $p_{01}^{(2)} = 0.25$, and  $\kappa=1.5$.}
	\label{fig:Pdv2}
\vspace{-0.4cm}
\end{figure}

In the second example, the parameters are taken as $\kappa=1.5$, $p_{10}^{(1)} = 0.3$,  $p_{10}^{(2)} = 0.2$, $p_{01}^{(1)} = 0.4$, and $p_{01}^{(2)} = 0.25$. In this case, there exist $8$ different deterministic decision rules in the form of \eqref{eq:bayesrule}, which achieve the pairwise error probability vectors marked with blue stars in Fig.~\ref{fig:Pdv2}. The minimum value of \eqref{eq:objectiveprospect} among these deterministic decision rules is $3.9278$, which corresponds to the pairwise error probability vector shown with the green square in the figure. In addition, the pairwise error probability vectors corresponding to the solutions with randomization of two and three deterministic decision rules are marked with the red triangle and the black circle, respectively. In this scenario, the minimum objective value ($3.8432$) can be achieved via randomization of two deterministic decision rules, as well. This is again in compliance with the theorem (case 2), which states that an optimal decision rule can be obtained as a randomization among \emph{at most} $M(M-1)+1$ deterministic decision rules of the form given in \eqref{eq:bayesrule}.

\section{Concluding Remarks}

This letter presents a unified characterization of optimal decision rules for simple $M-$ary hypothesis testing under a generic performance criterion that depends on arbitrary functions of error probabilities. It is shown that optimal performance with respect to the design criterion can be achieved by randomizing among at most two deterministic decision rules of the form  reminiscent (but not necessarily identical) to Bayes rule when points on the decision boundary do not contribute to the error probabilities. For the general case, the solution for an optimal decision rule is reduced to a search over two  weight coefficient vectors, each of length $M(M-1)$. Likelihood ratios are shown to be sufficient statistics. Classical performance measures including Bayesian, minimax, Neyman-Pearson, generalized Neyman-Pearson, restricted Bayesian, and prospect theory based approaches all appear as special cases of the considered framework.

Finally, we point out that the form of optimal local sensor decision rules for the problem of distributed detection \cite{Delic16,Neto17,Adve18,Warren99} with conditionally independent observations at the sensors and an \emph{arbitrary} fusion rule can be characterized using the proposed framework.

\begin{appendices}
\section{Proof of Lemma}\label{app:lemma}

Since $\{\bp\,:\, \bv^T \bp = \bv^T \bp_0\}$ is a supporting hyperplane to $\sfP$ at the point $\bp_0$, we get  $\bv^T \bp \geq  \bv^T \bp_0$ for all $\bp\in\sfP$. Furthermore, the deterministic decision rule given in \eqref{eq:bayesrule}, denoted here by $\bdelta^{\ast}$, minimizes $\bv^T \bp$ among all decision rules $\bdelta \in \sfD$ (and consequently over all $\bp\in\sfP$). Since $\bp_0\in \sfP$ as well, the deterministic decision rule given in \eqref{eq:bayesrule} achieves a performance score of $\bv^T \bp_0$. Any other decision rule that does not agree with $\bdelta^{\ast}$ on any subset of $\bar{\sfB}(\bv)$ with nonzero probability measure will have a strictly greater performance score than $\bv^T \bp_0$ (due to the optimality of $\bdelta^{\ast}$), and hence, cannot be on the supporting hyperplane.\\
\emph{Case 1}: We prove the first part by contrapositive. Suppose that the deterministic decision rule $\bdelta^{\ast}$ given in \eqref{eq:bayesrule} yields $\bp^{\ast}\neq \bp_0$ meaning that $\bp_0$ is achieved by some other decision rule $\bdelta^0 \in \sfD$. Since $\bdelta^{\ast}$ minimizes $\bv^T \bp$ over all $\bp\in\sfP$, $\bv^T \bp^{\ast} = \bv^T \bp_0$ holds and both $\bp^{\ast}$ and $\bp_0$ are located on the supporting hyperplane $\{\bp\,:\, \bv^T \bp = \bv^T \bp_0\}$. This implies that $\bdelta^{\ast}$  and $\bdelta^0$ must agree on any subset of $\bar{\sfB}(\bv)$ with nonzero probability measure. As a result, the difference between the pairwise probability vectors $\bp^{\ast}$ and $\bp_0$ must stem from the difference of $\bdelta^{\ast}$  and $\bdelta^0$ over $\sfB(\bv)$. Consequently, the set $\sfB(\bv)$ cannot have zero probability under all hypotheses.\\
\emph{Case 2}: Suppose that the set of boundary points specified by $\sfB(\bv)$ has nonzero probability under some hypotheses. In this case, each point in  $\sfB_{i,j}(\bv)$ can be assigned arbitrarily (or in a randomized manner) to hypotheses $\calH_i$ and $\calH_j$. Since the way the ties are broken does not change $\bv^T \bp$, the resulting error probability vectors are all located on  the intersection of the set $\sfP$ with the $M(M-1)-1$ dimensional supporting hyperplane $\{\bp\,:\, \bv^T \bp = \bv^T \bp_0\}$.  By Carath\'{e}odory's Theorem \cite{CONVEX_Rockefellar}, any point (including $\bp_0$) in the intersection set, whose dimension is at most  $M(M-1)-1$, can be represented as a convex combination of at most $M(M-1)$ extreme points of this set. Since these extreme points can only be obtained via deterministic decision rules which all agree with $\bdelta^{\ast}$ on the set $\bar{\sfB}(\bv)$, $\bp_0$ can be achieved by a randomization among at most $M(M-1)$ deterministic decision rules of the form given in \eqref{eq:bayesrule}, all corresponding to the weights specified by $\bv$.\hfill$\blacksquare$
\end{appendices}

\bibliographystyle{IEEEtran}
\bibliography{mybib}


\end{document}